\documentclass[5p,times,twocolumn]{elsarticle}
\usepackage{amsmath,amssymb}
\biboptions{comma,sort&compress}
\usepackage[utf8]{inputenc}
\usepackage{graphicx}
\usepackage{bm}
\usepackage[dvipsnames]{xcolor}
\usepackage{hyperref} 
\hypersetup{colorlinks=true, linkcolor=blue, anchorcolor=red, citecolor=blue, filecolor=red, urlcolor=red, pdfauthor=author}

\journal{Physics Letters B}

\begin{document}

\begin{frontmatter}

\title{Heavy quark polarization anisotropy as a novel probe of fireball geometry}

\author[niser]{Amaresh Jaiswal}
\ead{a.jaiswal@niser.ac.in}

\address[niser]{School of Physical Sciences, National Institute of Science Education and Research, An OCC of Homi Bhabha National Institute, Jatni-752050, India}

\date{\today}

\begin{abstract}
We propose a new approach to probe the initial fireball geometry in relativistic heavy-ion collisions using spin polarization. Specifically, we introduce polarization harmonics of open heavy hadrons as a novel observable sensitive to geometric anisotropies. Heavy quarks are produced in early hard scatterings and can acquire spin polarization from the strong, transient electromagnetic fields present at early times. As they propagate through the anisotropic quark-gluon plasma, medium-induced interactions lead to path-length dependent depolarization, imprinting an azimuthally anisotropic polarization pattern. Within the framework of rotational Brownian motion, we show that the resulting polarization harmonics are directly related to the initial spatial eccentricities, thereby establishing heavy-flavor polarization anisotropies as a sensitive and complementary probe of the early-time collision geometry. We present quantitative estimates of the second (elliptic) polarization harmonic associated with the recently observed $D^{*+}$ spin alignment reported by the ALICE Collaboration.
\end{abstract}

\end{frontmatter}

\section{Introduction}
\label{sec:intro}

The propagation of heavy quarks through the quark–gluon plasma (QGP) created in high-energy heavy-ion collisions provides a powerful and well-established probe of the medium’s transport properties~\cite{Moore:2004tg, Gubser:2006bz, Das:2010tj, Akamatsu:2008ge, Banerjee:2011ra, Ding:2012iy, vanHees:2007me, PhysRevD.37.2484, dong2019heavy, PhysRevC.73.034913, das2015toward, PhysRevLett.100.192301, ALICE:2021rxa, STAR:2017kkh}. A central goal of phenomenological studies of heavy-quark dynamics in the QGP is the extraction of transport coefficients, in particular the drag and momentum-diffusion coefficients that quantify the heavy quark's interaction with the surrounding medium. Most existing approaches concentrate on the translational Brownian motion of heavy quarks, in which medium-induced drag and diffusion govern the evolution of their linear momentum and spatial trajectories. In contrast, recent experimental observations of hadron polarization in relativistic heavy-ion collisions have opened a new avenue for exploring heavy-quark dynamics through their spin degrees of freedom~\cite{Liang:2004ph, Liang:2004xn, STAR:2017ckg, STAR:2018pps, STAR:2018fqv, Niida:2018hfw, STAR:2018gyt, ALICE:2019aid, STAR:2019erd, ALICE:2019onw, Singha:2020qns, Chen:2020pty, STAR:2020xbm, ALICE:2021pzu, STAR:2021beb, Mohanty:2021vbt, STAR:2022fan, STAR:2023eck, ALICE:2025cdf}. This has motivated recent theoretical efforts to investigate heavy-quark polarization within the framework of stochastic angular-momentum evolution and rotational Brownian motion~\cite{Dey:2025ail, Liu:2024hii, Li:2025ipk}.

Relativistic heavy-ion collisions create a short-lived, strongly interacting QGP~~\cite{Shuryak:2014zxa} whose macroscopic properties are governed not only by its microscopic dynamics but also by the geometry of the initial nuclear overlap region~\cite{Heinz:2013th}. Event-by-event fluctuations in the transverse distribution of matter give rise to characteristic spatial anisotropies that strongly influence the subsequent evolution of the system~\cite{Alver:2010gr}. In recent years, this sensitivity has enabled heavy-ion collisions to emerge as a novel tool for probing nuclear structure, with flow and geometry-driven observables providing access to deformation, shape fluctuations, and subnucleonic degrees of freedom of the colliding nuclei~\cite{Broniowski:2013dia, Giacalone:2019pca, Giacalone:2021udy, Jia:2022ozr, Jia:2025wey}. Quantifying the initial fireball geometry is therefore essential not only for constraining the transport properties and early-time dynamics of the QGP, but also for advancing our understanding of the structure of nuclei at high energies.

Flow harmonics have been extensively used to probe the fireball geometry in relativistic heavy-ion collisions, reflecting the conversion of initial-state spatial anisotropies into final-state momentum anisotropies through hydrodynamic expansion~\cite{Alver:2010gr, Heinz:2013th, Bhalerao:2020ulk}. Besides flow harmonics, the fireball geometry in relativistic heavy-ion collisions has also been probed through Hanburry-Brown-Twiss (HBT) femtoscopy~\cite{Baym:1997ce, Wiedemann:1999qn, Lisa:2005dd}, jet quenching~\cite{Bjorken:1982tu, Gyulassy:1990ye, Qin:2015srf}, heavy-flavor suppression~\cite{Svetitsky:1987gq, Rapp:2009my, Das:2015ana, Bhaduri:2018iwr, Bhaduri:2020lur}, and electromagnetic observables~\cite{Shuryak:1978ij, Kajantie:1986dh, Alam:1996fd}. While hydrodynamic flow and HBT are primarily sensitive to the collective expansion and freeze-out geometry of the medium, jet quenching, heavy-flavor suppression, and electromagnetic probes exhibit harmonic structures governed by path-length or source-geometry effects. Heavy-flavor polarization harmonics, introduced in this work, fall into this latter category, providing a new and complementary spin-based probe of the initial-state geometry.

Unlike light hadrons, which are predominantly emitted near the freeze-out hypersurface during hadronization, heavy quarks are produced almost exclusively in early-time hard partonic scatterings. As a result, they serve as clean and penetrating probes of the earliest stages of the hot, deconfined medium created in relativistic heavy-ion collisions. Moreover, in noncentral collisions, extremely strong but short-lived magnetic fields are generated, which are appreciable only during these early times. Such transient electromagnetic fields can induce spin polarization of heavy quarks along the magnetic-field direction. As polarized heavy quarks subsequently propagate through the anisotropic quark–gluon plasma, interactions with medium constituents lead to spin depolarization in a path-length–dependent manner~\cite{Dey:2025ail}. Consequently, heavy-quark polarization observables provide a unique and sensitive probe of the initial fireball geometry in relativistic heavy-ion collisions.

Recent studies have explored the role of heavy-quark angular momentum fluctuations in the early stages of relativistic heavy-ion collisions. In particular, Refs.~\cite{Pooja:2022ojj, Parisi:2025slf} investigated event-by-event fluctuations of heavy-quark orbital and spin angular momentum generated by anisotropic Glasma fields and demonstrated their sensitivity to the geometry of the initial state. The observable proposed in the present work is conceptually different. Rather than focusing on fluctuations of the initially generated angular momentum, we consider the in-medium evolution of an initially polarized heavy-quark ensemble and show that path-length-dependent spin depolarization in an anisotropic medium generates measurable polarization harmonics. The resulting observable is therefore sensitive to spin transport through the medium and can be extracted directly from event-plane-dependent measurements of heavy-hadron polarization, in direct analogy with conventional flow-harmonic analyses.

Unlike flow harmonics, which arise from collective momentum anisotropies generated by hydrodynamic expansion, the polarization harmonics proposed here originate from path-length-dependent spin depolarization during in-medium propagation. They therefore probe the initial geometry through an entirely different physical mechanism. In particular, the predicted signal is strongest at low and intermediate transverse momentum, decreases with increasing $p_T$, depends on the heavy-quark charge and mass, and differs qualitatively between mesons and baryons. These features provide characteristic signatures that distinguish polarization harmonics from conventional flow observables and from other possible sources of spin alignment.

In this Letter, we formulate heavy-flavor polarization harmonics as a new probe of the spatial anisotropy of the fireball created in relativistic heavy-ion collisions. Within the framework of rotational Brownian motion, we demonstrate that these polarization harmonics are directly linked to the initial-state spatial eccentricities, thereby establishing heavy-flavor polarization anisotropies as a sensitive and complementary probe of the early-time collision geometry. Unlike flow harmonics, polarization harmonics encode geometric information through spin depolarization during in-medium propagation rather than through collective hydrodynamic expansion. As an illustration, we provide quantitative estimates of the second (elliptic) polarization harmonic corresponding to the recently observed $D^{*+}$ meson spin alignment reported by the ALICE Collaboration. To the best of our knowledge, polarization harmonics have not previously been proposed as an observable in relativistic heavy-ion collisions. Throughout the text, three vectors are denoted in bold fonts and we use natural units where $c=\hbar=k_B=1$.


\section{Fireball geometry and path length}
\label{sec:geo_pl}

We consider heavy quarks produced at random points in the transverse plane and propagating along a fixed azimuthal direction $\phi$. The ensemble-averaged path length inside the medium, $\langle L(\phi)\rangle$, encodes information about the geometric anisotropies of the system and can be expressed in a form analogous to the harmonic expansion of anisotropic flow. Without loss of generality, the azimuthal dependence of the average path length can be written as a Fourier series,
\begin{equation}\label{eq:Lphi_general}
\langle L(\phi)\rangle = L_0 \left[ 1 + \sum_{n=2}^{\infty} 2\,\ell_n \cos n\! \left( \phi-\Psi_n \right) \right],
\end{equation}
where $L_0 \equiv \langle\langle L(\phi)\rangle\rangle_{\phi}$ is the azimuthally averaged path length, $\ell_n$ are the path-length anisotropy coefficients, and $\Psi_n$ denote the corresponding event-plane angles. The harmonics $n=2,3,4,\ldots$ correspond to elliptic, triangular, quadrangular, and higher-order geometric deformations of the medium.

For a circle of radius $R$, the mean chord length can be written as $\rho R$, where $\rho$ is a constant which depends on the statistical ensemble used to sample chords inside the circle~\cite{Santalo_2004}. Since each chord admits two possible directions of motion and a particle produced at a random point traverses only one side of the chord, the mean path length is exactly half the mean chord length, yielding $\langle L\rangle=\rho R/2$. We consider a weakly anisotropic transverse geometry whose boundary is parametrized as
\begin{equation}\label{eq:Rphi_app}
R(\varphi) = R_0 \left[ 1 + \sum_{n=2}^{\infty} a_n \cos n\! \left( \varphi-\Phi_n \right) \right]\!, \quad |a_n|\ll1 ,
\end{equation}
where $R_0$ is the mean radius, $a_n$ are the boundary anisotropies and $\Phi_n$ are the symmetry-plane angles, which are commonly approximated by the corresponding event-plane angles $\Psi_n$.

For straight-line propagation from production points distributed uniformly inside the medium, the ensemble-averaged path length for particles emitted at azimuthal angle $\phi$ scales with the effective transverse extent of the medium along that direction, i.e. $\langle L(\phi)\rangle\propto R(\varphi)$. For convex geometries with weak anisotropies, as considered in Eq.~\eqref{eq:Rphi_app}, this implies
\begin{equation}
\frac{\delta\langle L(\phi)\rangle}{L_0} \simeq \frac{\delta R(\phi)}{R_0},
\end{equation}
consistent with mean chord-length arguments in transport theory for convex media~\cite{Santalo_2004}. Here, $\delta\langle L(\phi)\rangle\equiv\langle L(\phi)\rangle-L_0$ and $\delta R(\phi)\equiv R(\phi)-R_0$. Comparing Eqs.~\eqref{eq:Lphi_general} and \eqref{eq:Rphi_app}, one immediately obtains
\begin{equation}\label{2lnan}
2\,\ell_n=a_n,
\end{equation}
which is a purely geometric relation valid for weak anisotropies up to linear order in the anisotropy.

The geometric anisotropies of the initial state in heavy-ion collisions are commonly characterized by~\cite{Bhalerao:2020ulk},
\begin{equation}\label{eq:epsn_def}
\epsilon_n \, e^{i n \Phi_n} \equiv - \frac{\langle r^n \, e^{i n\varphi}\rangle} {\langle r^n\rangle}, \qquad n \ge 2,
\end{equation}
where $\epsilon_n$ are the spatial eccentricities, $\Phi_n$ are the corresponding participant-plane angles, $(r,\varphi)$ are polar coordinates in the transverse plane and the averages are taken over the initial density profile. As customary, we use the event-plane angle $\Psi_n$ as a proxy for the participant-plane angle $\Phi_n$, since the latter cannot be accessed experimentally. In the weak-anisotropy limit, the path-length coefficients $\ell_n$ are linear in the corresponding spatial eccentricities, $\ell_n \propto \epsilon_n$, while higher harmonics can receive additional nonlinear contributions from lower-order eccentricities. To linear order in geometric deformations and for smooth density profiles, one finds (see the derivation in~\ref{app}),
\begin{equation}\label{eq:Lphi_small_eps}
\langle L(\phi)\rangle \simeq L_0 \left[ 1 - \sum_{n=2}^{\infty} \frac{\epsilon_n}{(n+2)} \, \cos n\! \left( \phi-\Psi_n \right) \right],
\end{equation}
where the minus sign reflects the fact that particles propagating along directions of larger transverse extent experience longer path lengths.


\section{Heavy quark polarization}
\label{sec:HQ_pol}

The stochastic Landau–Lifshitz–Gilbert (LLG) equation provides a natural framework for analyzing the evolution of heavy-quark spin polarization~\cite{Landau:1935qbc, Gilbert_2004, Nishino_2015, Meo_2023}. In the particle’s rest frame, the spin dynamics described by the LLG equation can be written as~\cite{Dey:2025ail}
\begin{equation}\label{SpinLangevin}
\frac{d\bm{s}}{d\tau} = \bm{s} \times \left[ \tilde{\bm{B}} + \boldsymbol{\xi}(\tau) \right] - \lambda\, \bm{s} \times \left( \bm{s} \times \tilde{\bm{B}} \right),
\end{equation}
where $\bm{s}$ denotes the classical spin vector and $\tau$ is the proper time. The effective field $\tilde{\bm{B}} \equiv \gamma\bm{B} = -\frac{\partial \mathcal{H}}{\partial \bm{s}}$ is defined in terms of the spin Hamiltonian $\mathcal{H}$, with 
$\gamma$ being the gyromagnetic ratio relating magnetic moment vector $\boldsymbol{\mu}$ and spin vector $\bm{s}$ via the relation $\boldsymbol{\mu} = \gamma\, \bm{s}$. The first term describes spin precession in the external magnetic field, while the double cross-product term provides Gilbert damping (with damping coefficient $\lambda$), driving the spin toward equilibrium without changing its magnitude. The fluctuating field $\boldsymbol{\xi}(\tau)$ follow the correlation properties of white noise, i.e., $\langle \xi_k(\tau) \rangle = 0$ and $\langle \xi_k(\tau_1)\, \xi_{l}(\tau_2) \rangle = 2\, D\, \delta_{kl} \,\delta(\tau_1 - \tau_2)$, where $D$ represents the spin diffusion coefficient.

We consider a sphere in spin space of fixed radius $s$, parametrized by $\bm{s}=(s,\theta,\phi_s)$, where each point corresponds to a distinct spin orientation of the particle~\cite{PhysRevE.76.051104, PhysRevA.11.280}. The probability for a spin-polarized particle to have an instantaneous orientation $(\theta,\phi_s)$ then defines a distribution on this sphere. Assuming an axially symmetric Hamiltonian and applying the Kramers–Moyal expansion to Eq.~\eqref{SpinLangevin} yields a Fokker–Planck equation governing the time evolution of this probability distribution $\mathcal{P}(\theta,\tau)$~\cite{Dey:2025ail}
\begin{equation}\label{FokPM}
\tau_s \frac{\partial \mathcal{P}}{\partial \tau} = 
\frac{1}{\sin \theta} \frac{\partial}{\partial \theta} \Bigg[ \sin \theta\, \Bigg(\frac{\partial}{\partial \theta} + \frac{\lambda}{D}\,\mu\, B(\tau) \sin \theta\, \Bigg) \Bigg]\,\mathcal{P}\,,
\end{equation}
where $\tau_s\equiv1/D$ represents the spin-relaxation timescale. For heavy quarks, the dominant effect of the strong initial magnetic field is encoded in the initial condition by assuming that the spins are aligned with the field direction, $\theta=\theta_{0}$. Specifically, we take $\mathcal{P}(\theta,0)=\frac{1}{2\pi}\delta(\cos\theta-\cos\theta_{0})$, which leads to the solution~\cite{Dey:2025ail}
\begin{align}\label{FP_sol}
\mathcal{P}(\theta,\tau;\theta_{0},0) = \sum_{k=0}^{\infty}\frac{2k+1}{4\pi}
&\,\exp\!\left[-k(k+1)\frac{\tau}{\tau_s}\right] \nonumber\\
& \times P_k(\cos\theta)\,P_k(\cos\theta_{0}) ,
\end{align}
where $P_k(z)$ are Legendre polynomials in $z$ of degree $k$. In obtaining the above result, magnetic field is neglected during the subsequent evolution, since it decays rapidly and its leading effect is already incorporated through the initial condition.

Using the solution for probability distribution in Eq.~\eqref{FP_sol}, one obtains~\cite{Dey:2025ail}
\begin{align}
\langle \cos\theta \rangle &= \cos\theta_{0} \, e^{- 2\tau/\tau_s}, \label{av_cos_th}\\
\langle \cos^2\theta \rangle &= \frac{1}{3} + \frac{2}{3} P_2( \cos\theta_{0})\, e^{- 6\tau/\tau_s}. \label{av_cos2_th}
\end{align}
Initially, heavy-quark spins align parallel ($\theta_0=0$) or antiparallel ($\theta_0=\pi$) to the magnetic field, depending on the quark charge. Accordingly, the polarization of open heavy baryons, which is proportional to $\langle \cos\theta \rangle$, depends on the heavy-quark charge through $\cos\theta_{0}=\pm1$. In contrast, the polarization of open heavy mesons is governed by $\langle \cos^{2}\theta \rangle$ and is therefore insensitive to the heavy-quark charge, since $P_{2}(\cos\theta_{0})=1$. The deviation of $\langle \cos\theta \rangle$ and $\left( \langle \cos^2\theta \rangle-1/3 \right)$ from zero quantify the polarization of open heavy baryons and mesons, respectively. Specifically, the spin polarization of baryons can be expressed as $|\bm{P}_B|=\frac{3}{\alpha_B}\langle \cos\theta \rangle$, where $\alpha_B$ is the baryon decay parameter. On the other hand, meson spin alignment is quantified by $\Delta\rho_{00}\equiv\rho_{00}-\frac{1}{3}=\frac{5}{2}\left( \langle \cos^2\theta \rangle-1/3 \right)$, where $\rho_{00}$ is element of spin density matrix. Consequently, the spin polarization/alignment can be written in the generic form as
\begin{equation}\label{pol_gen}
P \,(\bm{p}) = A\, e^{-\alpha\,\tau/\tau_s},
\end{equation}
where, $\bm{p}$ denotes the heavy-quark momentum, $A$ is a particle species dependent constant and $\alpha=2,\,6$ for baryons and mesons, respectively. 


\section{Polarization harmonics}
\label{sec:pol_har}

Accounting for the Lorentz contraction of $\langle L(\phi) \rangle$ in the heavy-quark rest frame, the duration over which the heavy quark undergoes Brownian motion in the QGP is $\tau = \left[ \langle L(\phi) \rangle\, m_{Q} \right]/|{\bm p}|$. Here $|{\bm p}|=\sqrt{p_{T}^{2}\cosh^{2}y+m_{Q}^{2}\,\sinh^{2}y}$, where $m_{Q}$ is the mass of the heavy-quark, $p_T$ is its transverse momentum and $y$ its rapidity. Substituting the expression for $\tau$ in Eq.~\eqref{pol_gen}, we obtain
\begin{equation}\label{pol_mid}
P\, (p_T,\,\phi,\,y) = A \exp \left(-\frac{\alpha\,m_Q\,\langle L(\phi) \rangle}{|{\bm p}|\,\tau_s} \right).
\end{equation}
Using Eq.~\eqref{eq:Lphi_small_eps} for $\langle L(\phi) \rangle$ and expanding to linear order in geometric anisotropies, we get
\begin{align}\label{pol_full}
P\, (p_T,\,\phi,\,y) = A &\exp \left(-\frac{\alpha\,m_Q\,L_0}{|{\bm p}|\,\tau_s} \right) \nonumber\\
&\times \left[ 1 + \sum_{n=2}^{\infty} 2\,p_n \cos n \left( \phi-\Psi_n \right) \right].
\end{align}
Here, we obtain the expression for polarization harmonics as 
\begin{equation}\label{pol_har}
p_n\,(p_T,y) = \frac{\alpha\,m_Q\,L_0\,\epsilon_n}{2\,(n+2)\,|{\bm p}|\,\tau_s},
\end{equation}
demonstrating that geometric anisotropies directly generate harmonic modulations in path-length dependent observables, resulting in the scaling relation $p_n\propto\epsilon_n$ at leading order in anisotropies. We note that the explanation of the $D^{*+}$ meson spin alignment proposed in Ref.~\cite{Dey:2025ail} corresponds to the first term on the right-hand side of Eq.~\eqref{pol_full}. Additionally, we introduce polarization harmonics as a new set of observables, which constitutes the central result of the present work.

At this juncture, it is important to clarify that Eq.~\eqref{pol_har} should be interpreted as a leading-order scaling relation that isolates the geometric origin of the polarization harmonics. In particular, within the weak-anisotropy and straight-line propagation limits, the harmonic coefficients inherit the same azimuthal structure as the initial-state eccentricities, i.e., $p_n\propto\epsilon_n$. The proportionality coefficient is model dependent and may be modified by realistic medium evolution, energy loss, and event-by-event fluctuations, but the linear scaling itself follows directly from the path-length dependence of spin depolarization. Further, the present analysis is carried out within a minimal setup that assumes a static medium with a constant average temperature and straight-line heavy-quark propagation. The purpose of these assumptions is not to provide a realistic description of the full heavy-ion evolution, but rather to isolate the essential mechanism by which an anisotropic path length converts an initially polarized heavy-quark ensemble into polarization harmonics. In this sense, the calculation plays a role analogous to early geometric estimates of jet-quenching anisotropies and electromagnetic probes, where the key physical scaling relation was identified before being embedded in more realistic dynamical simulations.

Since the polarization is exponentially sensitive to the path length, as obtained in Eq.~\eqref{pol_mid}, even modest azimuthal variations of $\langle L(\phi) \rangle$ generate a finite harmonic modulation. Subsequent medium evolution mainly rescales the effective path length and spin-relaxation time, and is therefore expected to modify the overall magnitude of $p_n$ rather than erase the harmonic structure itself. This is due to the fact that $p_n$ is fixed by the geometry of the anisotropic medium. Moreover, the experimentally accessible heavy-quark polarization is dominated by heavy quarks with moderate-to-large transverse momenta and hence relatively large transverse velocities~\cite{ALICE:2025cdf}. In contrast, the collective transverse flow of the medium develops gradually and remains smaller during the early stages relevant for spin depolarization. Consequently, heavy quarks are expected to experience the medium before substantial transverse expansion has occurred, so that medium expansion primarily modifies the overall magnitude of the polarization harmonics rather than their characteristic azimuthal structure.

Although the overall magnitude of the signal depends on the details of the medium evolution, the appearance of polarization harmonics is a robust consequence of three generic ingredients: an initial heavy-quark polarization, path-length-dependent depolarization, and an anisotropic medium geometry. Any mechanism satisfying these conditions will produce an azimuthally modulated polarization signal with the same harmonic structure as the underlying geometry. Other mechanisms, such as vorticity-induced polarization or hadronization effects, may also contribute to heavy-hadron spin alignment. However, these effects are expected to exhibit different dependencies on transverse momentum, particle species, and harmonic order. In particular, the mechanism proposed here predicts a suppression of the polarization harmonics with increasing $p_T$ and a characteristic scaling with the initial eccentricities. These features provide possible discriminants in future measurements.

\begin{figure}
    \centering
    \includegraphics[width=\linewidth]{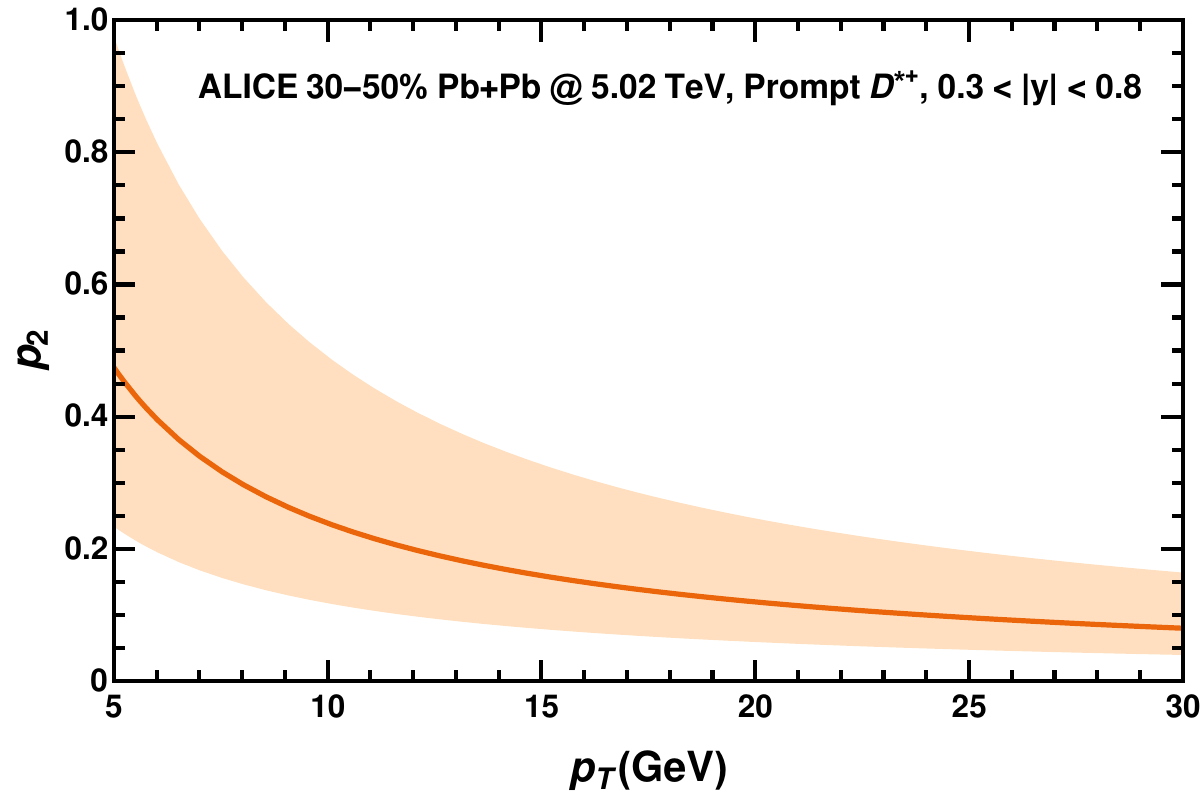}
    \caption{The elliptic polarization harmonic, $p_2$, as a function of transverse momentum $p_T$. The parameter values used in the plot are obtained from fits to ALICE measurements of the spin alignment of the $D^{*+}$ meson~\cite{ALICE:2025cdf, Dey:2025ail}. The shaded band is obtained by varying $L_0$ and $\epsilon_2$ by $\pm 20\%$ and $\tau_s$ by $\pm 30\%$.}
    \label{plot}
\end{figure}

To obtain quantitative estimates of the polarization harmonics, we use ALICE measurements of the spin alignment of the $D^{*+}$ meson in Pb$-$Pb collisions at $\sqrt{s_{\rm NN}}=5.02$~TeV, in the $30-50\%$ centrality class and $0.3<|y|<0.8$ rapidity interval~\cite{ALICE:2025cdf}. The observed increase of the $D^{*+}$ spin alignment with transverse momentum was interpreted in Ref.~\cite{Dey:2025ail} within a rotational Brownian motion framework. We adopt the corresponding fit parameters, $L_0=10$~fm and $\tau_s=1.31$~fm, and use $\alpha=6$ for mesons, charm-quark mass $m_Q=1.27$~GeV, and $y=0.55$ as the central value of the rapidity window. We consider $\epsilon_2=0.38$ for the elliptic geometric eccentricity from Monte Carlo Glauber calculations corresponding to $\sqrt{s_{\rm NN}}=5.02$~TeV Pb$-$Pb collisions in the $30-50\%$ centrality~\cite{Loizides:2017ack}. Figure~\ref{plot} shows the elliptic polarization harmonic $p_2$ as a function of the transverse momentum $p_T$. The shaded band is obtained by varying $L_0$ and $\epsilon_2$ by $\pm 20\%$ and $\tau_s$ by $\pm 30\%$ in order to account for uncertainties in the extraction of the model parameters. We observe that $p_2$ decreases with increasing $p_T$, reflecting the fact that high-$p_T$ charm quarks traverse the medium with minimal depolarization and are therefore less sensitive to the underlying geometric anisotropy.

To compute $p_T$-integrated polarization harmonics, an input for the heavy-quark momentum distribution is required. We employ a parametrized form of the prompt charm-quark spectrum obtained within the Fixed-Order plus Next-to-Leading Logarithm (FONLL) framework, which successfully reproduces the measured $D$-meson spectra in $pp$ collisions after fragmentation~\cite{Cacciari:1998it, Cacciari:2012ny, Cacciari:2015fta}:
\begin{equation}\label{param}
\left. \frac{dN}{d^2p_T\,dy} \right|_{\rm prompt} = \frac{a_0}{m_Q^2\left[ 1 + a_3 \left( \frac{p_T}{m_Q} \right)^{a_1} \right]^{a_2}}.
\end{equation}
The parameters are taken as $a_0=32.71558$, $a_1=1.95061$, $a_2=3.13695$, and $a_3=0.11981$, as obtained in Ref.~\cite{Liu:2019lac}. The momentum-differential polarization for charm quarks is then obtained by convoluting the heavy-quark momentum distribution in Eq.~\eqref{param} with the polarization function $P(p_T,\phi,y)$ in Eq.~\eqref{pol_full},
\begin{equation}
\frac{dP}{d^2p_T\,dy} = \left. \frac{dN}{d^2p_T\,dy} \right|_{\rm prompt} \times P\, (p_T,\,\phi,\,y).
\end{equation}
From the above equation, $p_T$ averaged elliptic polarization harmonic can be written as
\begin{equation}
\langle p_2 \rangle = \dfrac{\int p_T\,dp_T\,d\phi\, \cos\left[2( \phi-\Psi_2)\right] \frac{dP}{d^2p_T\,dy}}{\int p_T\,dp_T\,d\phi\, \frac{dP}{d^2p_T\,dy}}.
\end{equation}
Using the parameter values for the spin alignment of the $D^{*+}$ meson considered earlier~\cite{Dey:2025ail}, we obtain $\langle p_2 \rangle\simeq0.17\pm 0.07$.

It is important to note that the quantitative results provided above is only an order-of-magnitude estimate for the elliptic polarization harmonic. The numerical values obtained here should therefore be regarded as illustrative benchmarks rather than precision predictions. Nevertheless, the primary purpose is to demonstrate that the estimated magnitude is significant and the effect can naturally be of experimentally measurable magnitude. Experimentally, the proposed observable can be extracted from the event-plane dependence of the heavy-hadron polarization observable, e.g., $\rho_{00}(\phi-\Psi_n)$ for vector mesons or $P_H(\phi-\Psi_n)$ for heavy baryons. The corresponding harmonic coefficient may be obtained through a Fourier analysis with respect to the measured event plane, in direct analogy with conventional anisotropic-flow analyses. In practice, detector acceptance, finite event-plane resolution and possible decorrelation between participant-plane and event-plane angles will reduce the measured harmonic coefficients by the usual event-plane resolution factor, but will not modify the underlying characteristic harmonic structure. The present work does not claim that polarization harmonics are more sensitive than previously proposed angular-momentum observables in Ref.~\cite{Pooja:2022ojj, Parisi:2025slf}. Rather, they provide complementary information because they probe the initial magnetic field as well as integrated history of spin transport through the medium and are directly connected to experimentally measurable heavy-hadron polarization observables.


\section{Summary and outlook}
\label{sec:summary}

In this letter, we have introduced heavy-flavor polarization harmonics as a new class of observables sensitive to the spatial anisotropy of the fireball produced in relativistic heavy-ion collisions. Using a rotational Brownian motion framework, we established a direct connection between polarization harmonics and the initial-state spatial eccentricities, demonstrating that heavy-flavor polarization anisotropies provide a sensitive and complementary probe of early-time collision geometry. Unlike flow harmonics, which originate from collective momentum anisotropies generated by hydrodynamic expansion, polarization harmonics arise from path-length–dependent spin depolarization and therefore probe the initial geometry through an entirely different physical mechanism. We presented quantitative estimates of the second (elliptic) polarization harmonic associated with the recently observed $D^{*+}$ meson spin alignment reported by the ALICE Collaboration, illustrating the phenomenological relevance of the proposed observable. The proposed polarization harmonics establish spin as a new tomographic probe of the geometry of the quark–gluon plasma, complementary to flow, jet quenching, and femtoscopy.

The present analysis is formulated within a minimal framework that assumes a
static fireball characterized by a constant average temperature, allowing us
to isolate the essential physics underlying polarization harmonics.
Accordingly, the numerical values reported here are intended to provide a
qualitative, order-of-magnitude guidance and proof-of-principle study rather than precision predictions. A fully quantitative description will require embedding the spin dynamics in a realistic hydrodynamic background, including the time dependence of the temperature, magnetic field, heavy-quark momentum, and path length. Such effects are expected to modify the overall magnitude of the harmonics, but not the qualitative relation between polarization anisotropies and the initial geometry. In addition, a first-principles, field-theoretic determination of the heavy-quark spin relaxation time $\tau_s$ in the presence of external magnetic fields, as well as the inclusion of time-dependent heavy-quark energy loss from collisional and radiative processes in dynamical simulations, will be important for refining the quantitative extraction of the polarization harmonics $p_n$. These extensions are left for future work.

The observation of event-plane-dependent heavy-flavor polarization would open a qualitatively new window on the early stages of relativistic heavy-ion collisions, complementary to anisotropic flow and jet quenching, and would provide the first direct evidence that spin transport retains memory of the initial fireball geometry.

\section*{Acknowledgements}
%
The author thanks Partha Pratim Bhaduri, Santosh Kumar Das, Arun Kumar Yadav and Nachiketa Sarkar for helpful discussions. The author gratefully acknowledges Department of Atomic Energy (DAE), India for financial support.
%


\appendix 
\section{Relation between path-length harmonics and spatial eccentricities}
\label{app}

We consider a smooth, weakly deformed transverse geometry whose boundary can be parametrized as in Eq.~\eqref{eq:Rphi_app}. Retaining only a single harmonic $n$ for clarity, one has
\begin{equation}\label{eq:Rphi_def}
R(\varphi) = R_0\left[1+a_n\cos n(\varphi-\Psi_n)\right], \qquad |a_n|\ll1 ,
\end{equation}
Linearity ensures that the result generalizes straightforwardly to multiple harmonics. The spatial eccentricities are defined as
\begin{equation}\label{eq:epsn_def_app}
\epsilon_n e^{in\Psi_n} \equiv -\,\frac{\langle r^n e^{in\varphi}\rangle}{\langle r^n\rangle}, \qquad n\ge2 ,
\end{equation}
where the averages are taken over the transverse density. To linear order in the deformation it is sufficient to assume a uniform density inside the boundary.

The denominator in Eq.~\eqref{eq:epsn_def_app} is given by
\begin{equation}
\langle r^n\rangle = \int_0^{2\pi} d\varphi \int_0^{R(\varphi)} dr\, r^{n+1}.
\end{equation}
Performing the $r$ integration,
\begin{equation}
\langle r^n\rangle = \frac{1}{n+2} \int_0^{2\pi} d\varphi\, R^{n+2}(\varphi).
\end{equation}
Expanding Eq.~\eqref{eq:Rphi_def} to linear order,
\begin{equation}
R^{n+2}(\varphi) = R_0^{n+2} \left[ 1+(n+2)a_n\cos n(\varphi-\Psi_n) \right].
\end{equation}
Since the angular integral of the cosine vanishes,
\begin{equation}\label{eq:denominator_result}
\langle r^n\rangle = \frac{2\pi R_0^{n+2}}{n+2}.
\end{equation}

The numerator of Eq.~\eqref{eq:epsn_def_app} reads
\begin{equation}
\langle r^n e^{in\varphi}\rangle = \int_0^{2\pi} d\varphi\, e^{in\varphi} \int_0^{R(\varphi)} dr\, r^{n+1}.
\end{equation}
Performing the $r$ integration and expanding to linear order,
\begin{align}
\langle r^n e^{in\varphi}\rangle & = \frac{R_0^{n+2}}{n+2} \int_0^{2\pi} d\varphi\, e^{in\varphi} \nonumber\\
&\times\left[ 1+(n+2)a_n\cos n(\varphi-\Psi_n) \right].
\end{align}
The first term vanishes identically. Writing the cosine as
\begin{equation}
\cos n(\varphi-\Psi_n) = \frac{1}{2} \left[ e^{in(\varphi-\Psi_n)}+e^{-in(\varphi-\Psi_n)} \right],
\end{equation}
only the second exponential contributes to the angular integral, yielding
\begin{equation}
\int_0^{2\pi} d\varphi\, e^{in\varphi}\cos n(\varphi-\Psi_n) = \pi\,e^{in\Psi_n}.
\end{equation}
Thus,
\begin{equation}\label{eq:numerator_result}
\langle r^n e^{in\varphi}\rangle = \pi\, R_0^{n+2}\, a_n\, e^{in\Psi_n}.
\end{equation}

Taking the ratio of Eqs.~\eqref{eq:numerator_result} and \eqref{eq:denominator_result} and inserting it into the definition~\eqref{eq:epsn_def_app}, we obtain
\begin{equation}
\epsilon_n e^{in\Psi_n} = -\,\frac{(n+2)}{2}\,a_n\,e^{in\Psi_n},
\end{equation}
which immediately yields
\begin{equation}
\epsilon_n = -(n+2)\,a_n.
\end{equation}
The above relation is exact to linear order in the geometric deformation and holds for any smooth transverse profile. Using Eq.~\eqref{2lnan} in the above equation, we get
\begin{equation}
\ell_n = - \frac{\epsilon_n}{2\,(n+2)},
\end{equation}
providing a relation between path-length harmonics and spatial eccentricities.

\bibliographystyle{elsarticle-num}
\bibliography{ref}

\end{document}